\documentclass[aps,journal=pra,manuscript=article,twocolumn,superscriptaddress,floatfix]{revtex4}

\usepackage{appendix}
\usepackage{epsfig}
\usepackage{graphicx}
\usepackage{color}
\usepackage{dcolumn} 
\usepackage{bm} 
\usepackage{multirow} 
\usepackage{pifont} 
\usepackage{amsmath} 
\usepackage{subfigure}
\usepackage{float}
\usepackage{bm}
\begin{document}

\author{Daniel R. Nascimento}
\affiliation{
             Department of Chemistry and Biochemistry,
             Florida State University,
             Tallahassee, FL 32306-4390}

\author{A. Eugene DePrince III}
\affiliation{
             Department of Chemistry and Biochemistry,
             Florida State University,
             Tallahassee, FL 32306-4390}

\email{deprince@chem.fsu.edu}




\title{
Spatial and spin symmetry breaking in semidefinite-programming-based
Hartree-Fock theory
}



\begin{abstract}


The Hartree-Fock problem was recently recast as a semidefinite
optimization over the space of rank-constrained two-body reduced-density
matrices (RDMs) [Phys. Rev. A \textbf{89}, 010502(R) (2014)].  This
formulation of the problem transfers the non-convexity of the Hartree-Fock
energy functional to the rank constraint on the two-body RDM.  
We consider an equivalent optimization over the space of positive semidefinite
one-electron RDMs (1-RDMs) that retains the non-convexity of the
Hartree-Fock energy expression.  
The optimized 1-RDM satisfies ensemble
$N$-representability conditions, and ensemble spin-state conditions may be
imposed as well.  The spin-state conditions place additional linear and
nonlinear constraints on the 1-RDM.  We apply this RDM-based approach to
several molecular systems and explore its spatial (point group) and spin
($\hat{S}^2$ and $\hat{S}_3$) symmetry breaking properties. When imposing
$\hat{S}^2$ and $\hat{S}_3$ symmetry but relaxing point group symmetry,
the procedure often locates spatial-symmetry-broken solutions that are
difficult to identify using standard algorithms.  For example, the
RDM-based approach yields a smooth, spatial-symmetry-broken potential
energy curve for the well-known Be--H$_2$ insertion pathway.  We also
demonstrate numerically that, upon relaxation of $\hat{S}^2$ and
$\hat{S}_3$ symmetry constraints, the RDM-based approach is equivalent to
real-valued generalized Hartree-Fock theory.

\end{abstract}

\maketitle

\section{Introduction}

Hartree-Fock theory holds an important place in quantum chemistry. It
seldom provides a quantitative description of electronic structure, but it
serves as a useful starting point for more sophisticated electronic
structure methods, such as coupled-cluster theory
\cite{Cizek:1966:4256,Purvis:1982,Raghavachari:1989,Bartlett:2007:RMP:CC}.
The canonical form of the Hartree-Fock problem is that of Roothan
\cite{Roothan:1951:69} and Hall \cite{Hall:1951:541}, which involves the
repeated diagonalization of the Fock matrix.  When combined with
convergence acceleration procedures such as the direct inversion of the
iterative subspace (DIIS) \cite{Pulay:1980:393,Pulay:1982:556} and fast
two-electron repulsion integral generation (using, for example, graphical
processing units
\cite{Yasuda:2008:334,Ufimtsev:2008:222,Luehr:2011:949,Kalinowski:2017:3160}),
Hartree-Fock theory can be routinely applied to molecular systems
containing thousands of atoms \cite{Luehr:2011:949}.  Nevertheless, for
large enough systems, the diagonalization of the Fock matrix can
eventually become problematic and complicates the development of linearly
scaling algorithms.

The direct optimization of the one-electron reduced-density matrix (1-RDM) is an
attractive alternative to the iterative solution of the Roothan-Hall
equations for the molecular orbital coefficient matrix; this idea has been
widely explored since the 1950s
\cite{Lowdin:1955:1490,McWeeny:1956:496,McWeeny:1959:1528,McWeeny:1962:1028,McWeeny:1960:335}.
The most desirable feature of a density-matrix-based approach is that it
avoids the diagonalization of the Fock matrix, thereby facilitating the
development of linearly scaling algorithms
\cite{Mauri:1993:9973,Li:1993:10891,Stechel:1994:10088,Kussmann:2013:614}.
An immediate drawback, however, is that the 1-RDM associated with the
lowest possible energy does not correspond to any antisymmetrized
$N$-electron wave function, let alone one comprised of a single Slater
determinant.  To obtain a physically meaningful result, one must
explicitly consider the $N$-representability of the 1-RDM
\cite{Coleman:1963:668}.  Typically, $N$-representability (and
idempotency) in density-matrix-based Hartree-Fock is acheived through
``purification'' of an approximately $N$-representable 1-RDM
\cite{McWeeny:1960:335}.  Alternatively, recent work
\cite{Veeraraghavan:2014:010502,Veeraraghavan:2014:124107,Veeraraghavan:2015:022512}
has demonstrated the utility of semidefinite programming (SDP) techniques
for this problem.  The present work explores the SDP-based strategy.

Veeraraghavan and Mazziotti \cite{Veeraraghavan:2014:010502} recently
recast the Hartree-Fock problem as a constrained optimization over the
space of positive semidefinite two-body matrices.  The electronic energy
was expressed as a linear functional of the 1-RDM and a two-body matrix
(denoted ${}^2{\bf M}$ in Ref. \citenum{Veeraraghavan:2014:010502}) that
is related to the 1-RDM through a contraction.  By restricting the rank of
${}^2{\bf M}$, one can obtain a rigorous upper bound to the globally
optimal Hartree-Fock solution, and this ``rank-constrained
SDP''\cite{Veeraraghavan:2014:010502} optimization has the same formal
$\mathcal{O}(k^4)$ cost as the Roothan-Hall form of the problem.  Here,
$k$ represents the dimension of the one-electron basis set.
Alternatively, a lower bound can be obtained by imposing a relaxed set of
idempotency constraints on the full-rank matrix.  When separate
semidefinite optimizations under each set of constraints yield the same
1-RDM, that solution is guaranteed to be the globally optimal one.  This
guarantee is desirable; the price paid, though, is that the solution of
the lower-bound problem requires at least $\mathcal{O}(k^5)$
floating-point operations.



In this work, we consider an alternate representation of the
rank-constrained SDP problem that eliminates the consideration of the
two-body matrix, ${}^2{\bf M}$.  The result is a similar SDP algorithm
with a \textit{non-linear} objective function and \textit{non-linear}
constraints on the 1-RDM (in the case where the expectation value of
$\hat{S}^2$ is constrained).  The algorithm retains its formal
$\mathcal{O}(k^4)$ scaling, and, like other density-matrix-based
Hartree-Fock implementations, it avoids the repeated diagonalization of
the Fock matrix.  We demonstrate that the SDP-based approach can be
applied to several flavors of Hartree-Fock theory, including  restricted,
unrestricted, and generalized Hartree-Fock (RHF, UHF, and GHF,
respectively), depending on which spin symmetries are imposed on the
1-RDM.  We validate the implementation by exploring its spatial (point
group) and spin ($\hat{S}^2$ and $\hat{S}_3$) symmetry breaking properties
in several molecular systems.

\section{Theory}

\subsection{Density-matrix-based Hartree-Fock theory}

The electronic energy for the ground state of a many-electron system is a
function of the 1-RDM ($\bm{\gamma}$) and the two-electron reduced-density matrix (2-RDM,
$\bm\Gamma$)
\begin{equation}
E = \sum_{pqrs} \Gamma^{pr}_{qs} (pq|rs) + \sum_{pq} \gamma^p_q h_{pq}.
\end{equation}
Here, $(pq|rs)$ represents a two-electron repulsion integral in Mulliken
notation, $h_{pq}$ represents the sum of the one-electron kinetic
energy and electron-nuclear potential energy integrals, and the indices
$p$, $q$, $r$, and $s$ run over all spin orbitals.  The 1-RDM and 2-RDM are
defined as
\begin{eqnarray}
\gamma^p_q &=& \langle \Psi | \hat{a}_p^\dagger\hat{a}_q | \Psi \rangle, \\
\Gamma^{pr}_{qs} &=& \frac{1}{2} \langle \Psi | \hat{a}_p^\dagger\hat{a}_r^\dagger\hat{a}_s\hat{a}_q | \Psi \rangle,
\end{eqnarray}
where $\hat{a}^\dagger$ and $\hat{a}$ represent the fermionic creation and annihilation 
operators of second quantization, respectively.
Consider now the cumulant decomposition of the 2-RDM
\begin{equation}
{\bm \Gamma} = {\bm \gamma}\wedge{\bm \gamma} + {\bm \Delta_2},
\end{equation}
where the cumulant matrix, ${\bm \Delta}_2$, represents pure two-body
correlations, and the symbol $\wedge$ represents an antisymmetric tensor
product (or Grassman product) \cite{Coleman:1980:1279}.  By ignoring the
cumulant matrix, we arrive at a statistically-independent description of
electron motion; the resulting energy expression is equivalent to that
from Hartree-Fock theory
\begin{equation}
\label{EQN:en1}
E = \frac{1}{2} \sum_{pqrs} (\gamma^p_q\gamma^r_s - \gamma^p_s\gamma^r_q ) (pq|rs) + \sum_{pq} \gamma^p_q h_{pq}.
\end{equation}

To obtain the optimal Hartree-Fock 1-RDM, one can invoke the variational
principle and minimize the energy given by Eq. (\ref{EQN:en1}) with
respect to the elements of the 1-RDM.  Note, however, that this
optimization should be carried out under the constraint that the density
matrix be idempotent and have a trace equal to the number of electrons.
The idempotency condition is a specific manifestation of a more general
requirement that any physically meaningful density matrix should
correspond to an antisymmetrized $N$-electron wave function.  Necessary
ensemble $N$-representability conditions require that the eigenvalues of
the 1-RDM [the natural orbital (NO) occupation numbers] lie between zero and 
one and sum to the total number of electrons
\cite{Coleman:1963:668}.  The bounds on NO occupation numbers can be enforced
by requiring that the 1-RDM be positive semidefinite and related to
the one-hole density matrix, $\bar{\bm \gamma}$, by
\begin{equation}
\label{EQN:qcondition}
\gamma^p_q + \bar{\gamma}^q_p = \delta_{pq}.
\end{equation}
This one-hole density matrix, which must also be positive semidefinite,
is defined in second-quantized notation as
\begin{equation}
{\bar \gamma}^p_q = \langle \Psi | \hat{a}_p\hat{a}_q^\dagger | \Psi \rangle.
\end{equation}
The SDP procedure outlined in Sec. \ref{SEC:SDP} enforces these ensemble
$N$-representability conditions, rather than the idempotency condition.
It is important to note, however, that these conditions do not 
guarantee idempotency. Fortunately, as discussed in Sec.
\ref{SEC:CONCLUSIONS}, extensive numerical tests indicate that the
minimization of the electronic energy given by Eq. (\ref{EQN:en1}) with
respect to the elements of the 1-RDM under ensemble $N$-representability
constraints \textit{always} yields an idempotent 1-RDM.

For a non-relativistic Hamiltonian, the exact wave function should have a
well-definied total spin ($S$) and projection of spin ($M_S$).  Hence, one
can impose additional conditions on the 1-RDM that fix the particle number
and spin state for the system.  Particle number and $M_S$ can be fixed
according to
\begin{equation}
\label{EQN:tra}
{\rm Tr}({\bm \gamma}_{\alpha\alpha}) = N_\alpha,
\end{equation}
and
\begin{equation}
\label{EQN:trb}
{\rm Tr}({\bm \gamma}_{\beta\beta}) = N_\beta,
\end{equation}
where the subscripts $\alpha$ and $\beta$ refer to electrons of $\alpha$
and $\beta$ spin, and ${\bm \gamma}_{\alpha\alpha}$ and ${\bm
\gamma}_{\beta\beta}$ represent the spin-conserving blocks of the 1-RDM,
the full structure of which is
\begin{equation}
{\bm \gamma}=
 \begin{pmatrix}
  {\bm \gamma}_{\alpha\alpha}  & {\bm \gamma}_{\alpha\beta} \\
  {\bm \gamma}_{\beta\alpha}  & {\bm \gamma}_{\beta\beta} \\
 \end{pmatrix}.
\end{equation}
Note that the ${\bm \gamma}_{\alpha\beta}$ and ${\bm
\gamma}_{\beta\alpha}$ blocks are zero if the 1-RDM corresponds to a
wave function that is an eigenfunction of $\hat{S}_3$.  In this case, the
total spin quantum number is related to an off-diagonal trace of the
$\alpha\beta$--$\alpha\beta$ spin-block of the 2-RDM
\cite{Perez:1997:55,Gidofalvi:2005:052505}; for a
statistically-independent pair density, we have
\begin{equation}
\label{EQN:s2}
S(S+1) = \frac{1}{2} ( N_\alpha + N_\beta ) +  \frac{1}{4}( N_\alpha - N_\beta )^2 - {\rm Tr}({\bm \gamma}_{\alpha\alpha} {\bm \gamma}_{\beta\beta}).
\end{equation}
An ensemble $N$-representable 1-RDM corresponding to a state that is an
eigenfunction of $\hat{S}_3$ should satisfy Eqs.  (\ref{EQN:qcondition})
and (\ref{EQN:tra})-(\ref{EQN:trb}); for the 1-RDM to represent an ensemble
spin state with total spin, $S$, Eq.  (\ref{EQN:s2}) should also be
satisfied.

Various flavors of Hartree-Fock can be classified according to the
symmetries that are preserved by the wave function or density matrix
\cite{Fukutome:1981:955,Stuber:2003:67,Jimenez-Hoyos:2011:2667}.  
If we relax the spin
constraint given by Eq. (\ref{EQN:s2}), we arrive at 
density-matrix-based UHF.  If we also lift the constraint that the
Hartree-Fock wave function be an eigenfunction of $\hat{S}_3$,
we obtain GHF
\cite{Valatin:1961:1012,Kerman:1963:1326,Mestechkin:1974:45,Jimenez-Hoyos:2011:2667}.
In GHF, the particle number constraints of Eqs. (\ref{EQN:tra}) and
(\ref{EQN:trb}) reduce to a single constraint fixing the total particle
number
\begin{eqnarray}
    \label{EQN:trD}
    {\rm Tr}({\bm \gamma}_{\alpha\alpha}) + {\rm Tr}({\bm \gamma}_{\beta\beta}) = N
\end{eqnarray}
and ${\bm \gamma}$ and ${\bar{\bm{\gamma}}}$ are no longer guaranteed
to have a block-diagonal spin structure.  While $S$ may no longer be a good
quantum number, we can still evaluate the expectation value of $\hat{S}^2$ 
according to 
\begin{eqnarray}
\label{EQN:s2_2}
\langle \hat{S}^2 \rangle = \frac{3}{4} N + \frac{1}{4}[{\rm Tr}({\bm \gamma}_{\alpha\alpha})-{\rm Tr}({\bm \gamma}_{\beta\beta})]^2 \nonumber \\
-\frac{1}{4}{\rm Tr}({\bm \gamma}_{\alpha\alpha} {\bm \gamma}_{\alpha\alpha}) -\frac{1}{4} {\rm Tr}({\bm \gamma}_{\beta\beta} {\bm \gamma}_{\beta\beta}) \nonumber \\
+ \frac{1}{2}{\rm Tr}({\bm \gamma}_{\alpha\beta}){\rm Tr}({\bm \gamma}_{\beta\alpha}) + {\rm Tr}({\bm \gamma}_{\alpha\beta}{\bm \gamma}_{\beta\alpha}) - {\rm Tr}({\bm \gamma}_{\alpha\alpha} {\bm \gamma}_{\beta\beta}).
\end{eqnarray}

\subsection{Semidefinite optimization}

\label{SEC:SDP}

The minimization of the electronic energy given by Eq. (\ref{EQN:en1})
subject to the constraints outlined above constitutes a nonlinear
semidefinite optimization.  We adopt a matrix-factorization-based approach
to this problem based upon the ``RRSDP''
algorithm described in Refs. \citenum{Mazziotti:2004:213001} and \citenum{Mazziotti:2007:249}. The 1-RDM and
one-hole density matrix are expressed as contractions of auxiliary
matrices, ${\bm d}$ and $\bar{\bm d}$, as
\begin{equation}
\label{EQN:DECOMPOSITION}
\gamma^p_q = \sum_Q d^Q_p d^Q_q,
\end{equation}
and
\begin{equation}
\label{EQN:DECOMPOSITION_HOLE}
\bar{\gamma}^p_q = \sum_Q \bar{d}^Q_p \bar{d}^Q_q,
\end{equation}
and are thus positive semidefinite by construction. The
auxiliary matrices serve as the actual variable quantities in the
optimization.  Note that the spin structures of ${\bm \gamma}$ and
$\bar{\bm \gamma}$ depend on what spin symmetry is imposed. When
$\hat{S}_3$ symmetry is enforced, the density matrices consist of two
spin-conserving blocks which can be separately factorized as in Eqs.
(\ref{EQN:DECOMPOSITION}) and (\ref{EQN:DECOMPOSITION_HOLE}); in this
case, ${\bm d}$ and $\bar{\bm d}$ are also block diagonal, each comprised
of two $k \times k$ blocks, where $k$ represents the number of spatial
basis functions.  When $\hat{S}_3$ symmetry is broken, this block
structure is lost, and, like the density matrices, the auxiliary matrices
are comprised of a single $2 k \times 2 k$ block.

To obtain the optimal ${\bm d}$ and $\bar{\bm d}$, we minimize the 
augmented Lagrangian function
\begin{equation}
\label{EQN:L}
L({\bm d},\bar{\bm d}) = E({\bm d}) + \sum_i \bigg [ \frac{1}{\mu} C_i({\bm d},\bar{\bm d})^2 - \lambda_i C_i({\bm d},\bar{\bm d}) \bigg ],
\end{equation}
with respect to variations in their respective elements.  Here, the sum
runs over all constraints, $i$, the symbol, $C_i({\bm d},\bar{\bm d})$,
represents the error in constraint $i$, the symbol, $\lambda_i$,
represents the corresponding Lagrange multiplier, and $\mu$ is a penalty
parameter.  The optimization proceeds according to a two-step scheme that
is similar to that employed in Refs.  \citenum{Mazziotti:2004:213001}
and \citenum{Mazziotti:2007:249}:
\begin{enumerate}
\item{For a set of Lagrange multipliers 
$\{\lambda_i^{(n)}\}$ and penalty parameter $\mu^{(n)}$,
minimize Eq. (\ref{EQN:L}) with respect to the elements of ${\bm d}$ and
$\bar{\bm d}$.}
\item{Update the Lagrange multipliers 
\begin{equation}
\lambda_i^{(n+1)} = \lambda_i^{(n)} - 2
C_i^{(n)}({\bm d},\bar{\bm d})/\mu^{(n)} 
\end{equation}
and the penalty parameter 
\begin{equation}
\mu^{(n+1)} = f \mu^{(n)},
\end{equation} 
where $f$ is defined as
\begin{eqnarray}
    f= 
\begin{cases}
    1.0, & \text{if } \frac{\text{max }\{|C_i^{(n)}({\bm d})|\}}{\text{max }\{|C_i^{(n-1)}({\bm d})|\}} < 0.25 \\
    g, & \text{otherwise.}
\end{cases}
\end{eqnarray}
The parameter, $g$, is a random number that lies on the interval [0.08:0.12].  
}
\end{enumerate}
Steps 1 and 2 are repeated until the error in the constraints
($||{\bm C}({\bm d},\bar{\bm d})||$) falls below 10$^{-6}$ and the energy
changes between iterations by less than 10$^{-6}$ E$_{\rm h}$.  In a
typical optimization, there are fewer than 20 of these macroiterations.
The parameter, $g$, has been introduced into the RRSDP algorithm because
of the non-convex nature of the Hartree-Fock problem.  In cases where the
algorithm identifies a solution that is not the global solution, $g$
introduces some non-deterministic behavior to facilitate the identifaction
of additional solutions in subsequent computations.

In the present implementation, the minimization in Step 1 is achieved
using the Limited-memory
Broyden-Fletcher-Goldfarb-Shanno (L-BFGS) method, as implemented in the
library {\tt liblbfgs}.  The L-BFGS routine requires the repeated
evaluation of the electronic energy, the constraints, and the gradient of
the energy and constraints with respect to the elements of ${\bm d}$ and
$\bar{\bm d}$.
Evaluating the constraints and the
derivative of the constraints with respect to the elements of ${\bm d}$
and $\bar{\bm d}$ requires only $\mathcal{O}(k^2)$ floating-point operations.
The gradient of the energy with respect to the elements of
${\bm d}$ is given by the matrix product of ${\bm d}$ and the Fock matrix:
\begin{equation}
\frac{\partial E}{\partial d^Q_p} = 2 \sum_q d^Q_q F_{qp}
\end{equation}
Hence, the rate limiting step in the algorithm is the construction of the Fock
matrix, which requires $\mathcal{O}(k^4)$ floating-point operations.
Note that, as with the density matrices, the spin structure of the 
Fock matrix depends on whether or not $\hat{S}_3$ symmetry is enforced.  

The optimization is performed in the orthonormal basis defined by
L\"owdin's symmetric orthogonalization, and the initial ${\bm d}$ and
$\bar{\bm d}$ matrices are seeded with random numbers on the interval
[-1:1].  At the beginning of an optimization, the L-BFGS step can require
hundreds or thousands of Fock matrix builds, but this number
significantly decreases near convergence of the macroiterations.
Obviously, this semidefinite procedure requires that the Fock matrix be
built far more times than would be required by a conventional Hartree-Fock
algorithm.  However, it does have the nice property that the
diagonalization of the Fock matrix is avoided completely, unless, of
course, the orbital energies themselves are desired at the end of the
optimization.  Should one wish to devise a linear-scaling algorithm based
upon the formalism presented herein, it would be desirable to avoid the
diagonalization of the overlap matrix as well; in this case, the ensemble
$N$-representability conditions for the 1-RDM should be generalized
for non-orthogonal orbitals \cite{Veeraraghavan:2015:022512}.

\section{Results}

\label{SEC:RESULTS}

\subsection{Broken spatial symmetry}

Artifactual symmetry breaking problems often arise in Hartree-Fock-based
descriptions of strongly-correlated systems in which more than one
electronic configuration is important in the full configuration
interaction (CI) wave function.  Small {\em et al.}
\cite{Small:2015:024104} recently demonstrated that complex orbitals
resolve this issue for some well-known RHF-based cases.
\begin{figure*}[!htpb]
\caption{Potential energy curves for the C$_{\rm 2v}$
insertion of Be into H$_2$ in the cc-pVDZ basis set.  The curves are
shifted such that the energy of the colinear geometry is 0 kcal
mol$^{-1}$.  Results are provided at the (a) RHF [using conventional and
density-matrix-based (SDP-RHF) approaches] and (b) MP2 levels of theory.
MP2 results were obtained using orbitals from both approaches (indicated
in square brackets).}
\label{FIG:BEH2}
\includegraphics[scale=1.0]{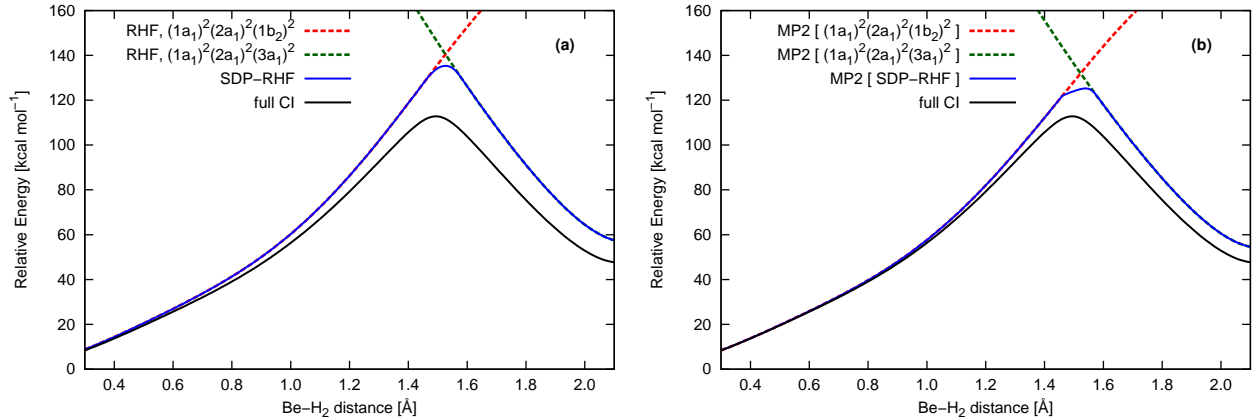}
\end{figure*}
An illustrative example considered in Ref. \citenum{Small:2015:024104} is
the potential energy curve (PEC) for the C$_{2\rm{v}}$ insertion of Be
into H$_2$, which has also served as a popular model for multireference
correlation studies
\cite{Purvis:1983:835,Mahapatra:1998:163,Mahapatra:1999:6171,Evangelista:2012:204108,Lyakh:2012:182,Brabec:2012:124102,Chen:2012:014108,Coe:2012:194111,Demel:2012:4753,Jimenez:2013:224110,Small:2015:024104}.
Figure \ref{FIG:BEH2}(a) illustrates the PEC as a function of the Be-H$_2$
distance, $x$, with the geometry (in units of \AA) defined as
\begin{eqnarray}
&\text { Be:}& \text{(0.0, 0.0, 0.0)}, \nonumber \\
&\text { H: }& \text{($x$,  1.344 - 0.46$x$, 0.0)}, \nonumber \\
&\text { H: }& \text{($x$, -1.344 + 0.46$x$, 0.0)}.
\end{eqnarray}
The curve labeled SDP-RHF was generated using the present
density-matrix-based algorithm while imposing $\hat{S}^2$ [Eq.
(\ref{EQN:s2})] and $\hat{S}_3$ symmetry [Eqs.
(\ref{EQN:tra})-(\ref{EQN:trb})].  Here, the one-electron basis set was
cc-pVDZ, and we employed the density fitting approximation to the
two-electron repulsion integrals (using the def2-TZVP-JK auxiliary basis
set).  Full CI computations employed conventional 4-index integrals.

As discussed in Ref. \citenum{Small:2015:024104}, the dominant
configuration in the full CI ground state is
(1a$_1$)$^2$(2a$_1$)$^2$(1b$_2$)$^2$ at $x = 0.0$ \AA~ (the colinear
geometry) and (1a$_1$)$^2$(2a$_1$)$^2$(3a$_1$)$^2$ at $x = 2.1$ \AA~
(separated Be + H$_2$).  Small {\em et al.} demonstrated that complex
orbitals yield a PEC that smoothly interpolates between the two RHF
limits, avoiding the cusp where these two configurations become
degenerate.  We see here that the present algorithm, which
employs real orbitals and density matrices, also yields a smooth PEC, and
inspection of the 1-RDM reveals that this smoothness is achieved through a
break in spatial symmetry.  Consider the point in the multireference
region of the PEC at $x=1.5$ \AA.  If we represent the optimized spin-free
1-RDM (${\bm \gamma}_{\alpha\alpha}+{\bm \gamma}_{\beta\beta}$) from the
present procedure in the basis of symmetry-pure orbitals obtained from
conventional RHF with the configuration,
(1a$_1$)$^2$(2a$_1$)$^2$(3a$_1$)$^2$, the first 4$\times$4 block has the
following structure:
\begin{eqnarray}
\begin{array}{lrrrrr} 
                      &~~~& \phi_{\rm 1a_1} &  \phi_{\rm 2a_1} &  \phi_{\rm 3a_1} &  \phi_{\rm 1b_2} \\ \\
      \phi_{\rm 1a_1} &~~~& 2.00  &     0.00  &     0.00   &    0.00 \\
      \phi_{\rm 2a_1} &~~~& 0.00  &     1.98  &    -0.13   &    0.08 \\
      \phi_{\rm 3a_1} &~~~& 0.00  &    -0.13  &     0.68   &    0.92 \\
      \phi_{\rm 1b_2} &~~~& 0.00  &     0.08  &     0.92   &    1.27 \\
\end{array}
\nonumber
\end{eqnarray}
There is significant mixing between orbitals of a$_1$ and b$_2$ symmetry,
which suggests that the symmetry of the overall wave function has been
reduced to at most C$_{\rm s}$.
Even when fully neglecting spatial
symmetry in a conventional RHF computation (using, for example, the implementation
in \textsc{Psi4} \cite{Psi4:1:1}), one in general locates either
one spatially-pure state or another, rather than the spatially
contaminated global solution.  

Figure \ref{FIG:BEH2}(b) illustrates the same curves generated using
second-order perturbation theory (MP2).  The MP2 computations built upon
orbitals from density-matrix-based Hartree-Fock (which are the
eigenfunctions of ${\bm \gamma}$) yield a PEC that 
interpolates between distinct RHF-based
MP2 curves with (1a$_1$)$^2$(2a$_1$)$^2$(1b$_2$)$^2$ and
(1a$_1$)$^2$(2a$_1$)$^2$(3a$_1$)$^2$ reference functions.  Here, however,
the curve in the multireference region is not smooth; there is a distinct
kink near $x=1.5$ \AA.  The broken symmetry MP2 curve also displays a
pronounced asymmetry in the multireference region; this behavior is quite
similar to that of MP2 built upon complex RHF orbitals, as presented in
Ref.  \citenum{Small:2015:024104}.

\begin{figure}[!htpb]
\caption{Potential energy curves for the dissociation of
N$_2$ in the cc-pVDZ basis set.$^a$ }
\label{FIG:N2}
\includegraphics[scale=1.0]{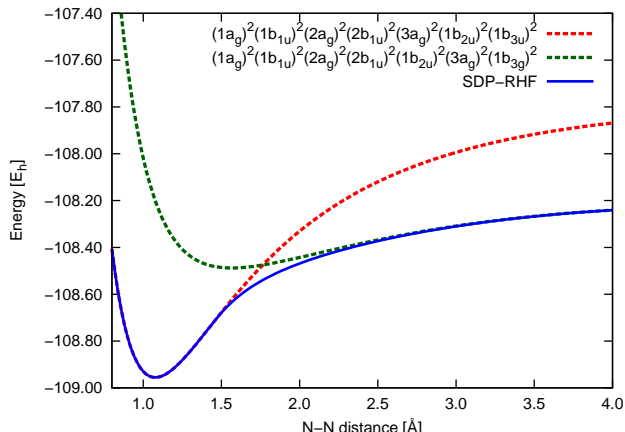}

\scriptsize{$^a$ RHF and SDP-RHF computations employ the density fitting
approximation and the cc-pVDZ-JK auxiliary basis set.}

\end{figure}

Veeraraghavan and Mazziotti \cite{Veeraraghavan:2014:010502} noted similar
symmetry-breaking behavior in other challenging multireference problems,
such as the dissociation of molecular nitrogen, within their rank-constrained
SDP
approach to Hartree-Fock theory.  This particular symmetry breaking
problem has been repeatedly studied using more conventional
Hartree-Fock algorithms as well
\cite{Xiangzhu:2009:085110,Thom:2008:193001,Small:2015:024104}.  Figure
\ref{FIG:N2} illustrates conventional RHF and density-matrix-based
Hartree-Fock PECs for the dissociation of N$_2$.  The RHF curves
correspond to configurations that differ only in the occupation of the
$\pi$ (1b$_{3\rm u}$ symmetry) and $\pi^*$ (1b$_{3\rm g}$ symmetry)
orbitals. As above, the PEC labeled SDP-RHF was generated using the
present approach while enforcing both $\hat{S}^2$ and $\hat{S}_3$
symmetry, and we observe the same spatial symmetry breaking behavior
reported elsewhere.
At equilibrium and dissociation, the density-matrix-based approach
yields solutions with energies that agree with those of one of the
spatially pure RHF configurations or the other.  
However, for N-N bond lengths in the range of 1.5--2.5 \AA, we 
obtain solutions that are lower in energy than both configurations; 
this stabilization is achieved 
via a break in the symmetry of the $\pi$/$\pi^*$ orbitals.

\subsection{Broken spin symmetry}

While it is desirable for the Hartree-Fock wave function to retain all of
the symmetries of the exact wave function, it is often useful, as in the
case of describing molecular dissociation, to lift constraints on its spin
symmetry ($\hat{S}^2$).  For example, it is well known that RHF in general
does not yield the correct dissociation limit for closed shell molecules;
UHF usually delivers size consistent results in such cases, at the expense
of retaining $S$ as a good quantum number.  Interestingly, even UHF does
not yield a size-consistent dissociation limit for some closed-shell
molecules, such as CO$_2$ \cite{Jimenez-Hoyos:2012:164109}.  If a truly
size-consistent single-determinant method is desired, one must be willing
to break additional symmetries in the wave function, such as $\hat{S}_3$.
In this Section, we explore the spin-symmetry breaking properties of
density-matrix-driven UHF and GHF (SDP-UHF and SDP-GHF, respectively) for
one well-studied \cite{Jimenez-Hoyos:2012:164109} case: the dissociation of
molecular oxygen.  The 1-RDMs generated from SDP-UHF satisfy ensemble
$N$-representability conditions and the $\hat{S}_3$ constraints of Eqs.
(\ref{EQN:tra})-(\ref{EQN:trb}), while those from SDP-GHF satisfy only
ensemble $N$-representability conditions.  For the remainder of this
Section, we use the terms UHF/SDP-UHF and GHF/SDP-GHF interchangeably.

Following Ref. \citenum{Jimenez-Hoyos:2012:164109}, the inability of UHF to yield a size-consistent dissociation curve for
O$_2$ is easily understood from simple spin arguments.  At equilibrium,
the ground state of molecular oxygen is the triplet state ($^3 \Sigma _{\rm
g}^-$), and the ground-state of the dissociation limit involves two
oxygen atoms in their triplet states ($^3$P).  The UHF wave function for a
given multiplicity is taken to be the high-spin determinant, so O$_2$ at
equilibrium and each dissociated oxygen atom all have $M_S = 1$.  At
dissociation, the two $M_S=1$ fragments can only couple to yield states of
$M_S=2$ (the quintet state) and $M_S=0$ (the singlet state).  Hence, the
UHF triplet state cannot connect the ground-state at equilibrium to those
(the singlet or quintet) at dissociation.

\begin{figure}[!htpb]
\caption{SDP-UHF and SDP-GHF potential energy curves for the dissociation
of O$_2$ in the cc-pVDZ basis set.$^a$ }
\label{FIG:O2}
\includegraphics[scale=1.0]{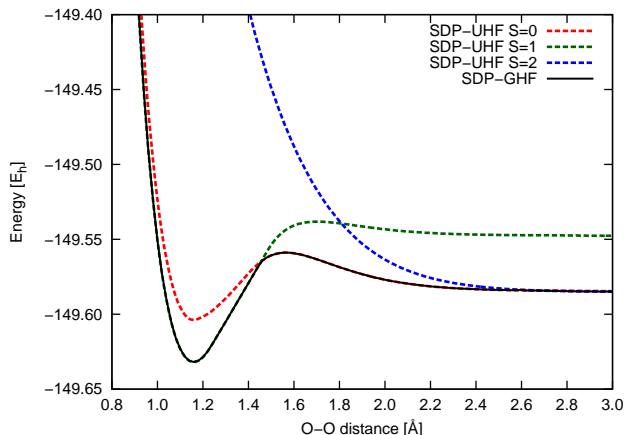}

\scriptsize{$^a$ SDP-UHF and SDP-GHF computations employ the density
fitting approximation and the cc-pVDZ-JK auxiliary basis set.}

\end{figure}

Figure \ref{FIG:O2} provides PECs for the dissociation of molecular oxygen
corresponding to the lowest-energy UHF singlet, triplet, and quintet
states, as well as that for GHF.  All computations were performed
within the cc-pVDZ basis set using the present density-matrix-based
approach to Hartree-Fock.
 As expected, the lowest-energy UHF curve at equilibrium
is that for the triplet state; the singlet UHF curve lies about 30
mE$_{\rm h}$ higher in energy. At dissociation, the singlet and quintet
curves become degenerate and are lowest in energy, while the dissociation
limit for the triplet lies about 40 mE$_{\rm h}$ higher in energy.  Hence,
as simple spin arguments suggest, the UHF triplet PEC does not connect the
lowest energy solutions at equilibrium and dissociation.  The present
results are in excellent agreement with those of Ref.
\citenum{Jimenez-Hoyos:2012:164109} and provide strong evidence for the
numerical equivalence between density-matrix-driven and conventional GHF.

\begin{figure}[!htpb]
\caption{The region of the potential energy curve for the dissociation of
O$_2$ where SDP-GHF provides a lower-energy solution than SDP-UHF (top
panel).  The value of $\langle \hat{S}^2 \rangle$ for SDP-GHF follows those
for the SDP-UHF triplet and singlet states at bond lengths less than 1.44
\AA~ and greater than 1.47 \AA, respectively (bottom panel).}
\label{FIG:O2_zoom}
\includegraphics[scale=1.0]{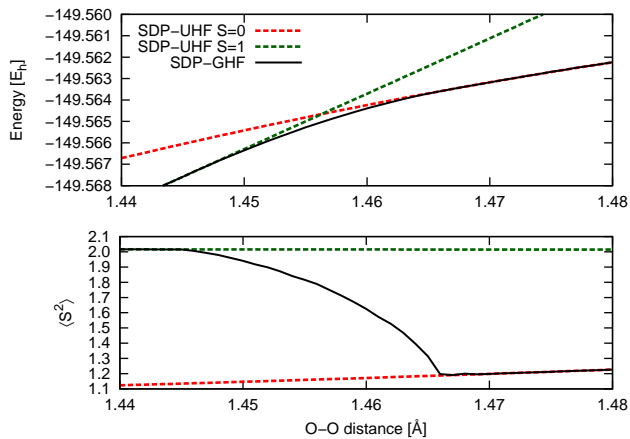}
\end{figure}
The UHF triplet and singlet states cross at an O--O distance slightly less
than 1.46 \AA.  The top panel of Fig. \ref{FIG:O2_zoom} provides a more
detailed illustration of the relevant PECs in the vicinity of this state
crossing.  We see that, in the region where the ground state predicted by
UHF changes from the triplet state to the singlet state, GHF offers a
solution that is lower in energy than either UHF solution.  As with the
PECs in Fig. \ref{FIG:O2}, this result is in excellent agreement with that
from Ref.  \citenum{Jimenez-Hoyos:2012:164109}.  The lower panel of Fig.
\ref{FIG:O2_zoom} illustrates how $\langle \hat{S}^2 \rangle$ changes as the GHF
PEC interpolates between the UHF triplet and singlet state curves.  We
note that while the qualitative behavior of $\langle \hat{S}^2 \rangle$ in this
region resembles that presented in Ref. \citenum{Jimenez-Hoyos:2012:164109},
there are quantitative differences between our computed values of $\langle
\hat{S}^2 \rangle$ and theirs.  For example, our computations indicate that the
expecation value of $\hat{S}^2$ merges with that of the UHF singlet curve 
at an O--O
distance less than 1.47 \AA, while the GHF and UHF values of $\langle \hat{S}^2
\rangle$ in Ref.  \cite{Jimenez-Hoyos:2012:164109} merge closer to 1.48
\AA.  Nonetheless, the results are quite similar.

\section{Conclusions}

\label{SEC:CONCLUSIONS}

We have considered the properties of a nonlinear representation of the
rank-constrained semidefinite programming
algorithm \cite{Veeraraghavan:2014:010502} for the Hartree-Fock problem.
The rate limiting step of the SDP-based approach is the formation of the
Fock matrix, so it shares the same formal $\mathcal{O}(k^4)$ scaling of
the usual Roothan-Hall formulation of the problem.  We demonstrated that
several flavors of Hartree-Fock theory (e.g. RHF, UHF, GHF) can be
implemented within this density-matrix-based formalism depending on which
spin symmetry conditions we choose to enforce, and we validated our
implementation by studying several spatial and spin symmetry breaking
problems.



Interestingly, each flavor of Hartree-Fock theory is recovered without
explicitly enforcing any constraints on the idempotency of the 1-RDM. We
only enforce ensemble $N$-representability conditions, which apply to both
correlated and uncorrelated 1-RDMs alike. This result contrasts with
conventional approaches to density-matrix-based Hartree-Fock theory which
rely on purification strategies to obtain idempotent density matrices.  We
note that we have verified numerically that the optimized 1-RDMs were
indeed idempotent in every computation performed in this work.  We can
rationalize the idempotency of our 1-RDMs by considering the structure of
the corresponding statistically independent pair density (${\bm \Gamma} =
{\bm \gamma}\wedge{\bm \gamma} $).  It can be show that the trace of such
a 2-RDM is minimized in the case of an idempotent 1-RDM and, accordingly,
an energy minimization procedure will favor idempotent 1-RDMs because they
minimize electron-electron repulsions.

Alternatively, the idempotency of the 1-RDMs can be understood from the
perspective of pure-state $N$-representability in RDM functional
theory \cite{Piris:2018:283}. In this context, Valone noted
\cite{Valone:1980:1344} that the distinction between pure-state and
ensemble $N$-representability is unnecessary in the case of the exact RDM
functional.  More broadly, pure-state $N$-representability can be achieved
under ensemble $N$-representability conditions, provided that the RDM
functional is an appropriate one (i.e. that the \textit{functional} is
pure-state $N$-representable).  Indeed, our observations are consistent
with this proposition. The Hartree-Fock energy functional is the RDM
functional that arises for a 1-RDM derived from a single Slater
determinant; the Hartree-Fock energy functional is thus pure-state
$N$-representable.  The pure-state $N$-representability of the functional
leads to additional desirable properties.  For example, the trace
constraint for GHF that defines the total particle number [Eq.
(\ref{EQN:trD})] technically only enforces the expectation value of $N$.
To specify the particle number exactly, the variance in $N$, $\langle \hat{N}^2
\rangle - \langle \hat{N} \rangle^2$, should vanish.  In the case that (i) the
2-RDM is expressed as an antisymmetrized product of the 1-RDM with itself
and (ii) the 1-RDM is idempotent, this variance is exactly zero.  As
discussed in the Appendix, similar arguments can be made for the exact
specification of the $M_S$ and $S(S+1)$. Hence, with the choice of the
Hartree-Fock energy functional as the RDM functional, the application of
ensemble $N$-representability and ensemble spin constraints yields
pure-state $N$- and $S$-representable 1-RDMs. \\

\appendix
\section{On the exact specification of $N$, $M_S$, and $S(S+1)$}


In the semidefinite-programming-based approach to Hartree-Fock theory, the
particle number and spin state for the system are specified through
constraints on the expectation values of $\hat{N}$, $\hat{S}_3$, and
$\hat{S}^2$.  The exact specification of these quantities technically
requires that the corresponding variances (e.g. $\langle \hat{N}^2 \rangle
- \langle \hat{N} \rangle^2$) be zero.  In the case that the 1-RDM is
idempotent and 2-RDM can be constructed as an antisymmetrized
product of the 1-RDM with itself, we can easily show that the variances
for $\hat{N}$ and $\hat{S}_3$ exactly vanish.  The
expectation value of the number operator squared leads to
\begin{equation}
\label{EQN:N2}
\langle \hat{N}^2 \rangle = \langle \hat{N} \rangle + \langle \hat{N} \rangle ^2 - {\rm Tr}({\bm \gamma}{\bm \gamma}),
\end{equation}
where we have used the fact that, at the Hartree-Fock level of theory,
${\bm \Gamma} = {\bm \gamma}\wedge{\bm \gamma}$, and we have assumed that trace
relations such as Eqs. (\ref{EQN:tra}) and (\ref{EQN:trb}) or Eq. (\ref{EQN:trD}) are satisfied.  If the 1-RDM is also
idempotent, the first and last terms in Eq. (\ref{EQN:N2}) cancel, and the
variance is zero.  The same analysis holds for number
operators corresponding electrons of $\alpha$ and $\beta$ spin individually
($\hat{N}_\alpha$ and $\hat{N}_\beta$, respectively).

The variance in $M_S$ is defined by $\langle \hat{S}_3^2 \rangle - \langle
\hat{S}_3 \rangle^2$, where $\hat{S}_3 = \frac{1}{2}(\hat{N}_\alpha -
\hat{N}_\beta)$, and $M_S = \langle \hat{S}_3 \rangle$, when Eqs.
(\ref{EQN:tra}) and (\ref{EQN:trb}) are satisfied.  
If the 2-RDM is expressible in terms of the 1-RDM, we have
\begin{eqnarray}
\label{EQN:MS2}
\langle \hat{S}_3^2 \rangle &=& \frac{1}{4} \bigg ( \langle \hat{N}_\alpha \rangle - {\rm Tr}({\bm \gamma}_\alpha{\bm \gamma}_\alpha) + \langle \hat{N}_\beta \rangle - {\rm Tr}({\bm \gamma}_\beta{\bm \gamma}_\beta) \nonumber \\
&+& \langle \hat{N}_\alpha\rangle^2 +  \langle \hat{N}_\beta\rangle^2 - 2  \langle \hat{N}_\alpha\rangle \langle \hat{N}_\beta\rangle
\bigg ).
\end{eqnarray}
In the case that 1-RDM is idempotent, Eq. (\ref{EQN:MS2}) reduces to
\begin{equation}
\label{EQN:MS2_2}
\langle \hat{S}_3^2 \rangle = \frac{1}{4} \bigg ( \langle \hat{N}_\alpha\rangle - \langle \hat{N}_\beta\rangle\bigg )^2 = \langle \hat{S}_3 \rangle^2,
\end{equation}
and the variance in $M_S$ vanishes.

\begin{figure}[!htpb]
\caption{The (a) total energy (E$_{\rm h}$), (b) expectation value of $\hat{S}^2$,
and (c) the Euclidean norm of the error in the pure spin conditions of Eq. (\ref{EQN:PURE_SPIN}) for 
the dissociation of molecular nitrogen in the cc-pVDZ basis set.$^a$}
\label{FIG:SPIN_N2}
\footnotesize
\input{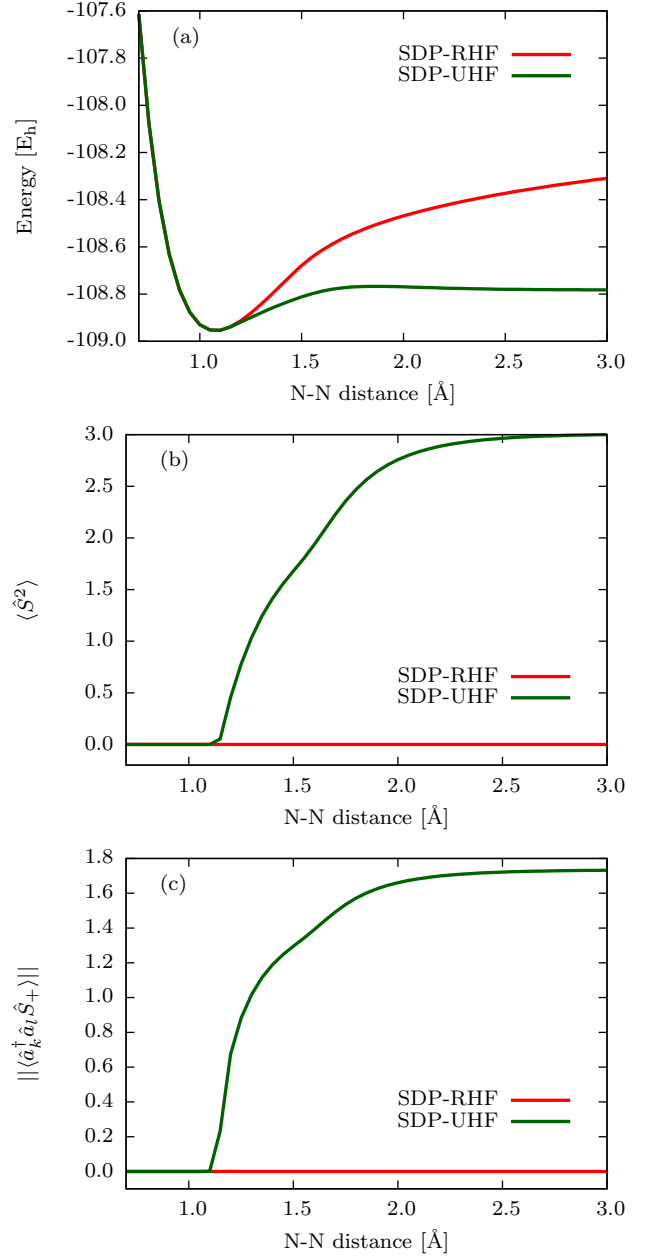}
\scriptsize{$^a$ SDP-RHF and SDP-UHF computations employ the density fitting
approximation and the cc-pVDZ-JK auxiliary basis set.}
\end{figure}

\begin{figure}[!htpb]
\caption{The (a) total energy (E$_{\rm h}$), (b) expectation value of $\hat{S}^2$,
and (c) the Euclidean norm of the error in the pure spin conditions of Eq. (\ref{EQN:PURE_SPIN}) for 
the dissociation of hydroxyl radical in the cc-pVDZ basis set.$^a$}
\label{FIG:SPIN_OH}
\footnotesize
\input{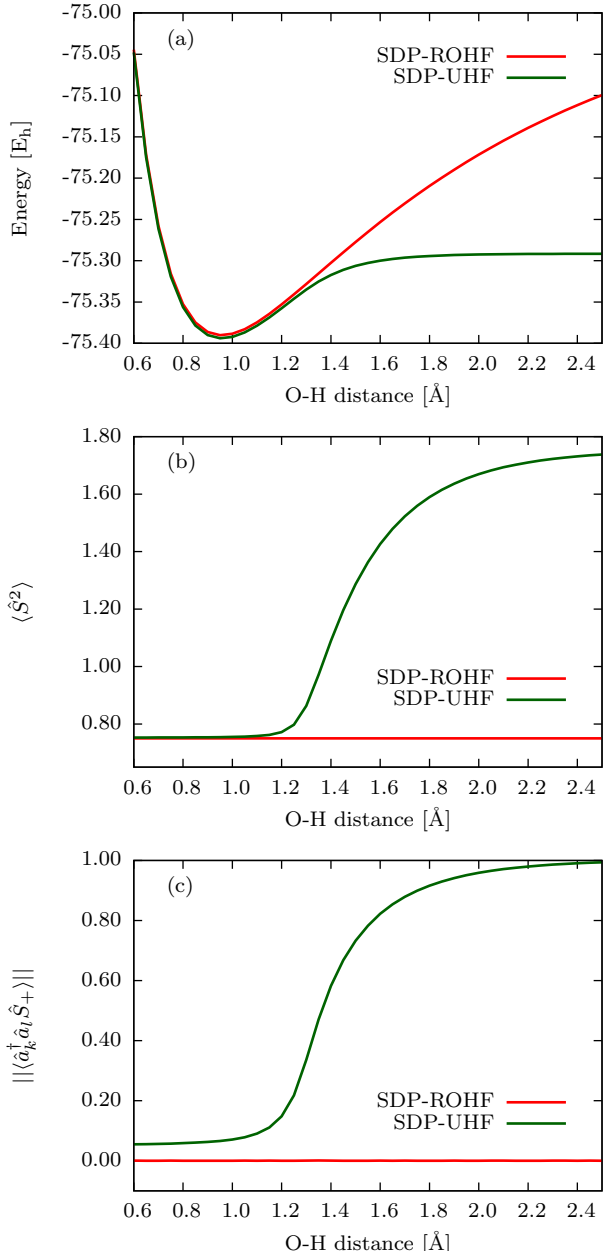}
\scriptsize{$^a$ SDP-ROHF and SDP-UHF computations employ the density fitting
approximation and the cc-pVDZ-JK auxiliary basis set.}
\end{figure}

A similar analysis of the variance in $S(S+1)$ is more involved, as the
expectation value of $\hat{S}^4$ depends upon the four-particle RDM.
Instead, we consider the two-particle pure-spin state conditions detailed
in Ref. \citenum{vanAggelen:136:2012}, the first of which is a
``contraction condition'' requiring that any pure spin state, $|\Psi^{SM_S}\rangle$,
should satisfy
\begin{equation}
\langle \Psi^{SM_S} | \hat{a}^\dagger_k \hat{a}_l (N\hat{S}_3 - M_S \hat{N})|\Psi^{SM_S}\rangle = 0,
\end{equation}
for all spin orbitals, $k$ and $l$.  As with the variances above, this condition is
automatically satisfied in the case that the 2-RDM is represented as an
antisymmetrized product of an idempotent 1-RDM with itself and that
Eqs. (\ref{EQN:tra}) and (\ref{EQN:trb}) are satisfied.  A less trival
constraint requires that the maximal spin projection for a spin pure state
satisfies
\begin{equation}
\langle \Psi^{SM_S} | \hat{a}^\dagger_k \hat{a}_l \hat{S}_+|\Psi^{SM_S}\rangle = 0,
\end{equation}
for all spin orbitals, $k$ and $l$.  For a 1-RDM with
$\hat{S}_3$ symmetry and a 2-RDM expressible in terms of the 1-RDM, this
set of constraints reduces to
\begin{equation}
\label{EQN:PURE_SPIN}
\forall k,l: \gamma^{k_\beta}_{l_\beta} - \sum_p \gamma^{p_\alpha}_{l_\alpha}\gamma^{k_\beta}_{p_\beta} = 0,
\end{equation}
where the Greek subscripts denote the spin component of the orbital label.
These constraints may be satisfied for a singlet state, where 
${\bm \gamma}_{\alpha\alpha} = {\bm \gamma}_{\beta\beta}$, and the 1-RDM
is idempotent.  However, these constraints are not obviously satisfied for
other spin states.  Here, we
demonstrate numerically that these constraints are satisfied in SDP-based
Hartree-Fock through
constraints on the expectation values of $\hat{S}_3$ and $\hat{S}^2$.

%

Figure \ref{FIG:SPIN_N2}(a) illustrates the potential energy curve for the
dissociation of molecular nitrogen at the SDP-RHF and SDP-UHF levels of
theory.  The expectation values of $\hat{S}^2$ from RHF and UHF [Fig.
\ref{FIG:SPIN_N2}(b)] diverge as the respective energies diverge, beyond
the Coulson-Fischer point.  Figure \ref{FIG:SPIN_N2}(c) shows the norm of
the constraints given by Eq. (\ref{EQN:PURE_SPIN}); we can see that these
constraints are satisfied by SDP-RHF (the norm is zero), whereas the norm
computed from the SDP-UHF 1-RDM becomes quite large beyond the
Coulson-Fischer point.

Lastly, we consider a similar analysis for a non-singlet case: the
dissociation of the hydroxyl radical.  Figure \ref{FIG:SPIN_OH}
provides results obtained using SDP-based restricted open-shell
Hartree-Fock (SDP-ROHF) and SDP-UHF.  The 1-RDM obtained from SDP-ROHF
satisfies Eqs. (\ref{EQN:tra}) and (\ref{EQN:trb}) as well as the
$\hat{S}^2$ constraint of Eq. (\ref{EQN:s2}) with $S$=0.5.  Panels (b) and
(c) of Fig. \ref{FIG:SPIN_OH} demonstrate that SDP-ROHF yields the correct
value of $S(S+1)$ at all O-H distances, and the norm of the errors defined
by Eq. (\ref{EQN:PURE_SPIN}) are zero for the entire curve as well.  The
corresponding data for SDP-UHF show clear deviations from the respective
values for a pure spin state, as expected.\\

\noindent {\bf Acknowledgments:}\\

This work was supported as part of the Center for Actinide Science and
Technology (CAST), an Energy Frontier Research Center funded by the U.S.
Department of Energy, Office of Science, Basic Energy Sciences under Award
No. DE-SC0016568.

\bibliography{scf}

\end{document}